\documentclass[journal=jctcce, manuscript=article,layout = twocolumn]{achemso}
\setkeys{acs}{articletitle=true}

\usepackage[version=3]{mhchem} 
\usepackage[T1]{fontenc}

\usepackage{booktabs}
\usepackage{color}
\usepackage{graphicx}
\usepackage{dcolumn}
\usepackage{bm}
\usepackage{dcolumn}
\usepackage{gensymb}
\usepackage{siunitx}
\usepackage{etoolbox}
\usepackage{booktabs}
\usepackage{multirow}
\usepackage[table,xcdraw]{xcolor}
\usepackage{tabularx}
\usepackage{makecell}
\usepackage{lipsum}
\usepackage{amsmath}
\usepackage{footnote}
\makesavenoteenv{table}
\newcommand{\RNum}[1]{\uppercase\expandafter{\romannumeral #1\relax}}

 \usepackage{adjustbox}
 \usepackage{breqn}
 \usepackage{array,multirow}
  
\title{Torsional potentials of glyoxal, oxalyl halides and their thiocarbonyl derivatives: Challenges for popular density functional approximations. }
\date{\today}

\author{Diana N. Tahchieva}
\affiliation{Institute of Physical Chemistry and National Center for Computational Design and Discovery of Novel Materials (MARVEL), Department of Chemistry, University of Basel, Klingelbergstrasse 80, CH-4056 Basel, Switzerland}

\author{Dirk Bakowies}
\affiliation{Institute of Physical Chemistry and National Center for Computational Design and Discovery of Novel Materials (MARVEL), Department of Chemistry, University of Basel, Klingelbergstrasse 80, CH-4056 Basel, Switzerland}

\author{Raghunathan Ramakrishnan}
\affiliation{Institute of Physical Chemistry and National Center for Computational Design and Discovery of Novel Materials (MARVEL), Department of Chemistry, University of Basel, Klingelbergstrasse 80, CH-4056 Basel, Switzerland}

\author{O. Anatole von Lilienfeld}
\email{anatole.vonlilienfeld@unibas.ch}
\affiliation{Institute of Physical Chemistry and National Center for Computational Design and Discovery of Novel Materials (MARVEL), Department of Chemistry, University of Basel, Klingelbergstrasse 80, CH-4056 Basel, Switzerland}

\begin{document}
\begin{abstract} 

The reliability of popular density functionals was studied for the description of torsional profiles of 36 molecules: glyoxal, oxalyl halides and their thiocarbonyl derivatives. HF and \textcolor{black}{eighteen} functionals of varying complexity, from local density to range-separated hybrid approximations and double-hybrid, have been considered and benchmarked against CCSD(T)-level rotational profiles. For molecules containing heavy halogens, all functionals except M05-2X and M06-2X fail to reproduce barrier heights accurately and a number of functionals introduce spurious minima. Dispersion corrections show no improvement. Calibrated torsion-corrected atom-centered potentials rectify the shortcomings of PBE and also improve on $\sigma$-hole based intermolecular binding in dimers and crystals.
 
\end{abstract}

\maketitle
%
\section{Introduction}

\begin{figure}[t]
\centering
\includegraphics[scale=0.3]{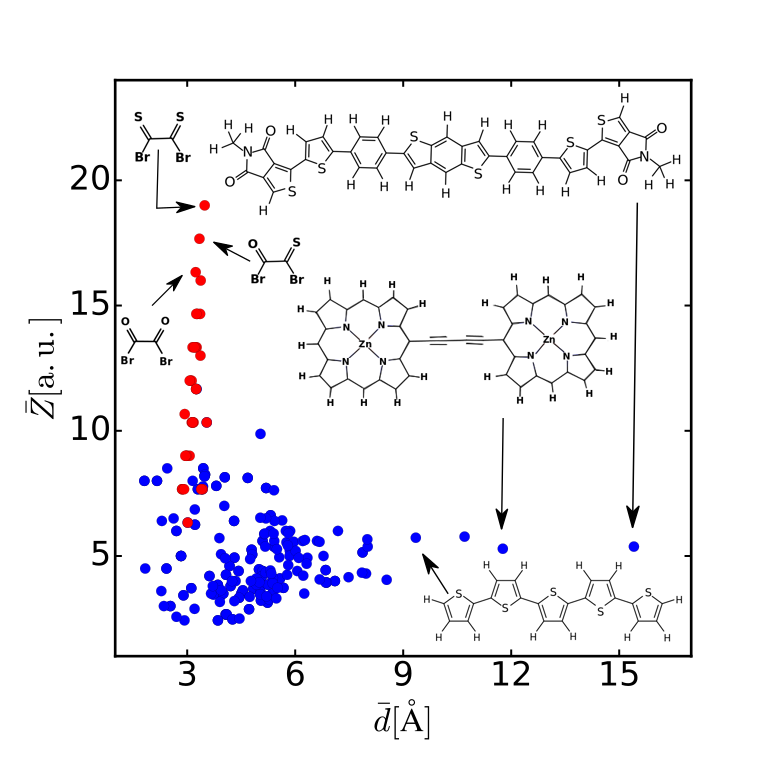}
\captionof{figure}[]{\label{hist_trend} Chemical diversity of torsional potentials in molecules studied in the literature 
\cite{imamura1968electronic,bixon1967potential,gorenstein1977torsional,kveseth1978conformational,
allinger1980torsional,van1982torsional,laskowski1985conformational,steele1985ab,darsey1987ab,carpenter1988analysis,caminati1991microwave,bock1991theoretical,
dixon1992torsional,guo1992ab,pelz1993torsional,koput1993torsional,hassett1993torsional,stanton1993ab,mannfors1994torsional,rothlisberger1994performance,hernandez1994ab,orti1995ab,koput1995ab,klopper1995ab,danielson1995conformational,koput1996ab,rothlisberger1996torsional,viruela1997geometric,karpfen1997single,ivanov1997torsional,senent1998ab,viruela1998difficulties,millefiori1998theoretical,graf1998ab,cui1998intermolecular,karpfen1999torsional,tsuzuki1999torsional,bongini1999theoretical,rablen1999hyperconjugation,tsuzuki2000high,koput2000equilibrium,bell2000far,de2000structure,sancho2001description,sancho2001torsional,arulmozhiraja2001torsional,federsel2001structure,shishkov2001trifluoromethoxy,tsuji2001molecular,sancho2001high,senent2001theoretical,sakata2001ab,sancho2001characterizing,sancho2002theoretical,kapustin2002torsion,tonmunphean2002torsional,orlandi2002torsional,pan2002theoretical,raos2003computational,duarte2003ab,klocker2003structure,klocker2003surprisingly,klocker2003trends,fliegl2003ab,cinacchi2003dft,sakata2003ab,halpern2003intrinsic,karpfen2004accurate,csaszar2004ab,miller2004quantum,senent2004ab,sancho2005torsional,perez2005torsional,puzzarini2005ab,kim2005extremely,goodman20051,klauda2005ab,zhou2005fc,mourik2005potential,fabiano2006torsional,hafezi2007study,haworth2007evaluating,chowdary2007vibrationally,sturdy2007torsional,tsuzuki2008conformational,ovsyannikov2008theoretical,de2008torsional,dorofeeva2008molecular,aquilanti2009intramolecular,tuttolomondo2009new,newby2009jet,koroleva2009internal,durig2010conformational,sidir2010ab,beames2011analysis,friesen2011structure,villa2011ccsd,castro2011analysis,blancafort2011quantum,lima2011thermodynamic,alyar2011torsional,hamza2012electronic,lee2012effect,qu2013full,senent2013highly,dakkouri2013theoretical,napolion2013accurate,bloom2014benchmark,planells2014effect,krkeglewski2014inversion,alparone2014structural,lin2015theoretical,kannengiesser2015acetyl,prashanth2016investigation,prashanth2016molecular,zakharenko2016torsion,deb2017theoretical,shachar2017conformational,lin2017torsional,goerigk2017look}(blue) and in this work (red). Vertical and horizontal axes correspond to averaged atomic charges and inter-atomic distances, respectively. Select examples are shown as insets.
}
\end{figure}

Rotational barriers play a crucial role in the dynamical mechanism of chemical reactions and processes, and influence a plethora of molecular properties and phenomena including fluorescence emission intensity \cite{srujana2016fluorescence}, intersystem crossing \cite{salaneck1989thermochromism} and protein folding \cite{sorokina2016role}. A quantitative description of torsional profiles in conjugated systems is often difficult to achieve in an experiment, but high-level \textit{ab-initio} methods can provide significant insight \cite{fabiano2006torsional,chen2011first,mo2007theoretical,bulat2002theoretical,karpfen1999torsional,viruela1997geometric}. It is particularly important that these numerical approaches yield accurate inter-atomic potentials, as chemical reactions often exhibit complex transition pathways with multiple local minima; notable examples being Diels-Alder reactions, which are often highly regio- and stereoselective \cite{rivero2017computational}. 
\\Coupled-cluster theory has emerged as a highly accurate method and is often used as a reference for benchmarking more approximate approaches. However, its computational complexity limits its use to relatively small molecules. Density functional theory (DFT), on the other hand, often offers good computational efficiency and reliability. Various DFT functionals have in fact shown promising performance for torsional profiles, but the inclusion of exact exchange appears to be important for $\pi$-conjugated systems \cite{sancho2001description,chen2011first}. 
\\An overview of the existing literature indicates a severe lack of chemical diversity in systems studied so far (Fig. \ref{hist_trend}). Most research thus far has been limited to first and second row elements in the periodic table \cite{imamura1968electronic,bixon1967potential,gorenstein1977torsional,kveseth1978conformational,
allinger1980torsional,van1982torsional,laskowski1985conformational,steele1985ab,darsey1987ab,carpenter1988analysis,caminati1991microwave,bock1991theoretical,
dixon1992torsional,guo1992ab,pelz1993torsional,koput1993torsional,hassett1993torsional,stanton1993ab,mannfors1994torsional,rothlisberger1994performance,hernandez1994ab,orti1995ab,koput1995ab,klopper1995ab,danielson1995conformational,koput1996ab,rothlisberger1996torsional,viruela1997geometric,karpfen1997single,ivanov1997torsional,senent1998ab,viruela1998difficulties,millefiori1998theoretical,graf1998ab,cui1998intermolecular,karpfen1999torsional,tsuzuki1999torsional,bongini1999theoretical,rablen1999hyperconjugation,tsuzuki2000high,koput2000equilibrium,bell2000far,de2000structure,sancho2001description,sancho2001torsional,arulmozhiraja2001torsional,federsel2001structure,shishkov2001trifluoromethoxy,tsuji2001molecular,sancho2001high,senent2001theoretical,sakata2001ab,sancho2001characterizing,sancho2002theoretical,kapustin2002torsion,tonmunphean2002torsional,orlandi2002torsional,pan2002theoretical,raos2003computational,duarte2003ab,klocker2003structure,klocker2003surprisingly,klocker2003trends,fliegl2003ab,cinacchi2003dft,sakata2003ab,halpern2003intrinsic,karpfen2004accurate,csaszar2004ab,miller2004quantum,senent2004ab,sancho2005torsional,perez2005torsional,puzzarini2005ab,kim2005extremely,goodman20051,klauda2005ab,zhou2005fc,mourik2005potential,fabiano2006torsional,hafezi2007study,haworth2007evaluating,chowdary2007vibrationally,sturdy2007torsional,tsuzuki2008conformational,ovsyannikov2008theoretical,de2008torsional,dorofeeva2008molecular,aquilanti2009intramolecular,tuttolomondo2009new,newby2009jet,koroleva2009internal,durig2010conformational,sidir2010ab,beames2011analysis,friesen2011structure,villa2011ccsd,castro2011analysis,blancafort2011quantum,lima2011thermodynamic,alyar2011torsional,hamza2012electronic,lee2012effect,qu2013full,senent2013highly,dakkouri2013theoretical,napolion2013accurate,bloom2014benchmark,planells2014effect,krkeglewski2014inversion,lin2015theoretical,kannengiesser2015acetyl,prashanth2016investigation,prashanth2016molecular,zakharenko2016torsion,deb2017theoretical,shachar2017conformational,lin2017torsional,goerigk2017look} and only a few studies considered molecules with other atom types, such as Br\cite{karpfen2004accurate,sidir2010ab}, Ge \cite{klocker2003surprisingly}, As \cite{klocker2003surprisingly}, Se \cite{bloom2014benchmark} and Te \cite{alparone2014structural} (see Figure \ref{hist_trend}). 
\\In this work, we assess the accuracy of a large range of popular density functionals for the description of single bond torsions in glyoxal, oxalyl and thiocarbonyl halides. Our focus is on extending compositional diversity, including heavier halogens (up to Br) in a conjugated carbonyl and thiocarbonyl scaffold (Fig. \ref{hist_trend} in red). The effect of dispersion corrections on torsional profiles is studied as well as custom-tailored atom centered potentials. The latter are shown to also improve the description of $\sigma$-hole based intermolecular binding.

\section{Computational details}

Torsional energy profiles $E(\Theta)$ were obtained through restricted geometry optimizations in which the torsional angle $\Theta = \Theta_{X\textrm{CC}\textit{Y}} =\Theta_{A\textrm{CC}\textit{B}}$ ($A, B$: oxygen or sulfur, $X, Y$: hydrogen or halogen) is kept constant and carbon atoms remain in the plane of their three bonded neighbors. The entire range of 0$^{\circ} < \Theta < 180^{\circ}$ was scanned in steps of $\Delta \Theta$ = 20$^{\circ}$. Note that $E(360^{\circ} - \Theta) = E(\Theta$), follows from the applied constraints. Calculations were carried out with Gaussian09 \cite{g09}, using Hartree-Fock as well as the following density functional approximations: LDA (SVWN5\cite{slater1951simplification,vosko1980accurate}), GGA (PW91\cite{perdew1991electronic}, PBE \cite{perdew1996generalized}, BLYP \cite{becke1988density,lee1988development}, BP86 \cite{becke1988density,perdew1986density}), mGGA (TPSS \cite{tao2003climbing}, M06L \cite{zhao2006new}), hGGA (PBE0 \cite{adamo1999toward}, B3LYP \cite{becke1988density,lee1988development,stephens1994ab}), mhGGA (M05 \cite{zhao2005exchange}, M06 \cite{zhao2008m06}, M05-2X \cite{zhao2006design}, M06-2X \cite{zhao2008m06}, M06-HF \cite{zhao2008m06}), RS (CAM-B3LYP \cite{yanai2004new}, M11 \cite{peverati2011improving}), double hybrid (B2PLYP \cite{grimme2006semiempirical}, \textcolor{black}{DSD-PBEP86(B3BJ) \cite{kozuch2011dsd}). Note that the DSD-PBEP86 functional has been parameterized for use with D3BJ correction and the expected accuracy of the double hybrid for non-bonding interactions cannot be obtained without the dispersion add-on \cite{kozuch2011dsd}.} The def2QZVPP \cite{weigend2005balanced} basis set was used throughout and parametric dispersion corrections (D3 \cite{grimme2010consistent}) were used in some cases. 
\\Additional plane-wave (PW) calculations were carried out for the PBE and BLYP functionals, using VASP \cite{kresse1996efficient,kresse1999ultrasoft} and CPMD \cite{hutter2005car}, respectively. MBD (Many Body Dispersion\cite{tkatchenko2012accurate}, for PBE) and DCACP \cite{von2004optimization,lin2007library} (Dispersion Corrected Atom-Centered Potential, for BLYP) corrections were used as indicated in Section 3.
MBD calculations were carried out using VASP in a box with size 14$\times$14$\times$10\: \AA$^3$\: (1$\times$1$\times$1 $\Gamma$-centered k-point grid) and a cutoff of 600\:eV. DCACP energies were calculated using a unit cell of (14\:\AA$)^3$\: and a cutoff of 200\:Ry with isolated boundary conditions.
\\Torsion corrected atom-centered potentials (TCACP) were constructed for PBE in analogy to other ACPs, following previously introduced optimization procedures \cite{von2005variational,lilienfeld2013force}. The TCACP is added to Goedecker's norm-conserving potentials \cite{goedecker1996separable} and shares the analytical form of its non-local part  (Eq. \ref{potential}). 
\begin{dmath}
V_I^{\rm{TCACP}}(r,r') = 
\sum^{+l}_{m=-l} Y_{lm}(\hat{\textbf{r}}) p_{l}(r)h^{l}_{11}p_{l}(r') Y^*_{lm}(\hat{\textbf{r}}'),
\label{potential}
\end{dmath}
with normalized projector $p_l(r) \propto r^l \rm{exp}[\frac{-r^2}{2r_{\textit{l}}^2}] $. $l$ denotes angular momentum (here $l=3$), $\hat{\textbf{r}}$ is the unit vector in the direction of $\textbf{r}$,  $r$=|\textbf{r}$-$ \textbf{R}$_I$| is the distance from the position of the nuclus $I$ and $Y_{lm}$ is a spherical harmonic. The parameters $h^{l=3}_{11}$ and $r_l$ are generated by minimizing a penalty function $P$, which has the following form:

\begin{equation}
P = \frac{1}{N}\sum_i^N |E_i^{\rm{CCSD(T)}} -  E^{\rm{PBE+TCACP}}_i|, 
\label{penalty}
\end{equation}
where $i$ runs over all $N=90$ conformations used in the training set. The penalty is minimized using the Nelder-Mead simplex-downhill algorithm \cite{lilienfeld2013force, von2004optimization, nelder1965simplex}, optimizing parameters corresponding to distance \cite{goedecker1996separable} $r_{l=3}$ from the position of the nuclei and amplitude\cite{goedecker1996separable} $h^{l=3}_{11}$. Parameterization of TCACPs was carried out with CPMD using a box size of (14\:\AA$)^3$\: and a 150\:Ry cutoff with isolated boundary conditions.
\\Binding energies with TCACPs were calculated using a cutoff of 150\:Ry and cell dimensions of 24$\times$17$\times$14\:\AA$^3$\: for oxalyl bromide-water and of 26$\times$19$\times$14\:\AA$^3$\: for the oxalyl bromide dimer. Cohesive energies as a function of a lattice scan were calculated for the oxalyl bromide crystal structure (2 molecules/unit cell) using Quantum Espresso \cite{giannozzi2009quantum} (PBE, PBE+TCACP) and VASP (PBE+MBD) with a $3\times3\times3$ $\Gamma$-centered k-point grid and a cutoff of 200\:Ry and 600\:eV, respectively. Experimental data was used for the crystal structure geometry and the initial unit cell dimensions, which was subsequently multiplied by a scaling factor $f$, ranging from 0.85 to 1.5.  
\\CCSD(T)  \cite{purvis1982full,raghavachari1989fifth} energies were calculated using Molpro \cite{werner2002molpro} and correlation-consistent basis sets \cite{dunning1989gaussian,woon1993gaussian} for M05-2X optimized geometries. CCSD(T)-F12\cite{adler2007simple} energies using cc-pVTZ-F12 \cite{peterson2008systematically} basis sets (including effective core potentials (ECP) for Br \cite{peterson2003systematically}) were calculated with Molpro \cite{werner2015molpro} for the same M05-2X geometries.

\section{Results and Discussion}
\subsection{CCSD(T) convergence test}

CCSD(T), often considered to be the gold standard of quantum chemistry, has been chosen as a reference level to judge the quality of density functional calculations. A basis set convergence analysis has been performed for  a few representative cases with different shapes of torsional potentials (O$_2$C$_2$Br$_2$, OSC$_2$HBr and S$_2$C$_2$Br$_2$).

\begin{table}[t]
\caption{CCSD(T) basis set convergence: Potential energy difference of oxalyl bromide between dihedral angles $\Theta$ = 80$^{\circ}$ and $\Theta$ = 180$^{\circ}$ (in kcal/mol). Yellow colouring shows the reference method used in this study.}

\begin{adjustbox}{width=\linewidth}
\begin{tabular}{ c c c| c c c  }
\multicolumn{3}{c}{\LARGE{Scheme A}} & \multicolumn{3}{c}{\LARGE{Scheme B}}\\
 \multicolumn{3}{c}{\LARGE{Valence \textit{sp} electrons correlated}} & \multicolumn{3}{c}{\LARGE{Valence \textit{sp} and 3\textit{d} (Br) electrons correlated}}\\\hline
\rule{0pt}{5ex}
\LARGE{ X }& \LARGE{ cc-pVXZ }& \LARGE{ aug-cc-pVXZ} & \LARGE{X} & \LARGE{  cc-pVXZ} & \LARGE{ aug-cc-pVXZ } \\\hline
\multicolumn{6}{c}{} \\
\multicolumn{6}{c}{\huge{OSC$_2$HBr (class I)}} \\\hline
\rule{0pt}{5ex}
\LARGE{D} & \huge{4.30}  & \huge{3.46} & \huge{D} & \huge{4.34} & \huge{3.46} \\
\rule{0pt}{5ex}
\LARGE{T}& \huge{3.62}  & \colorbox{yellow}{\huge{3.58}} &\huge{T} & \huge{3.68}  & \huge{3.72} \\
\rule{0pt}{5ex}
\LARGE{Q} & \huge{3.56}  &  \huge{3.55} & \huge{Q} & \huge{3.56}  & \huge{3.86} \\ 
\rule{0pt}{5ex}
\LARGE{5}& \huge{3.53}  & \huge{3.46} & \huge{5} & \huge{3.60}  & \huge{3.74}  \\
\rule{0pt}{5ex}
\LARGE{CBS (Q5)} & \huge{3.57} & \huge{3.63} & \LARGE{CBS (Q5)} & \huge{3.60} & \huge{3.63} \\
\rule{0pt}{5ex}
\begin{tabular}{@{}c@{}} \LARGE{CCSD(T)-F12} \\ \LARGE{(VTZ-F12); Ansatz 3C(FIX)} \end{tabular}  & \huge{3.52} & \huge{-} & \begin{tabular}{@{}c@{}} \LARGE{CCSD(T)-F12} \\ \LARGE{(VTZ-F12); Ansatz 3C(FIX)} \end{tabular}  & \huge{3.51} & \huge{-} 
\\\hline
& & & & & \\ 
\multicolumn{6}{c}{\huge{O$_2$C$_2$Br$_2$  (class II)}} \\\hline
\rule{0pt}{5ex}
\LARGE{D} & \huge{1.86}  & \huge{0.41} & \huge{D} & \huge{1.73} & \huge{0.41} \\
\rule{0pt}{5ex}
\LARGE{T}& \huge{0.61}  & \colorbox{yellow} {\huge{0.53}} &\huge{T} & \huge{0.61}  & \huge{0.62} \\
\rule{0pt}{5ex}
\LARGE{Q} & \huge{0.49}  &  \huge{0.54} & \huge{Q} & \huge{0.52}  & \huge{0.81} \\ 
\rule{0pt}{5ex}
\LARGE{5}& \huge{0.47}  & \huge{0.55} & \huge{5} &  \huge{0.47}  & \huge{0.71}  \\
\rule{0pt}{5ex}
\LARGE{CBS (Q5)} & \huge{0.51} & \huge{0.56} & \LARGE{CBS (Q5)} & \huge{0.47} & \huge{0.58} 
\\	
\rule{0pt}{5ex}
\begin{tabular}{@{}c@{}} \LARGE{CCSD(T)-F12} \\ \LARGE{(VTZ-F12); Ansatz 3C(FIX)} \end{tabular}  &  \huge{0.47}  &  \huge{-}  & \begin{tabular}{@{}c@{}} \LARGE{CCSD(T)-F12} \\ \LARGE{(VTZ-F12); Ansatz 3C(FIX)} \end{tabular}  & \huge{0.46} &  \huge{-} \\\hline

\multicolumn{6}{c}{} \\
\multicolumn{6}{c}{\huge{S$_2$C$_2$Br$_2$ (class III)}} \\\hline
\rule{0pt}{5ex}
\LARGE{D} & \huge{-3.05}  & \huge{-3.23} & \huge{D} & \huge{-2.98} & \huge{-3.19} \\
\rule{0pt}{5ex}
\LARGE{T}& \huge{-3.13}  & \colorbox{yellow} {\huge{-2.62}} &\huge{T} & \huge{-3.11}  & \huge{-2.55} \\
\rule{0pt}{5ex}
\LARGE{Q} & \huge{-2.57}  &  \huge{-2.32} & \huge{Q} & \huge{-2.48}  & \huge{-2.09} \\ 
\rule{0pt}{5ex}
\LARGE{5}& \huge{-2.40}  & \huge{-2.26} & \huge{5} &\huge{-2.36}  & \huge{-2.11}  \\
\rule{0pt}{5ex}
\LARGE{CBS (Q5)} & \huge{-2.19} &  \huge{-2.18} & \LARGE{CBS (Q5)} & \huge{-2.21} & \huge{-2.11} 
\\
\rule{0pt}{5ex}
\begin{tabular}{@{}c@{}} \LARGE{CCSD(T)-F12} \\ \LARGE{(VTZ-F12); Ansatz 3C(FIX)} \end{tabular}  &  \huge{-2.34}  &  \huge{-}  & \begin{tabular}{@{}c@{}} \LARGE{CCSD(T)-F12} \\ \LARGE{(VTZ-F12); Ansatz 3C(FIX)} \end{tabular}  & \huge{-2.35} &  \huge{-} 
\\\hline
\end{tabular}
\end{adjustbox}
\label{ccsdt_converge}
\end{table}

\begin{figure}[tp]
\centering
\includegraphics[scale=0.26]
{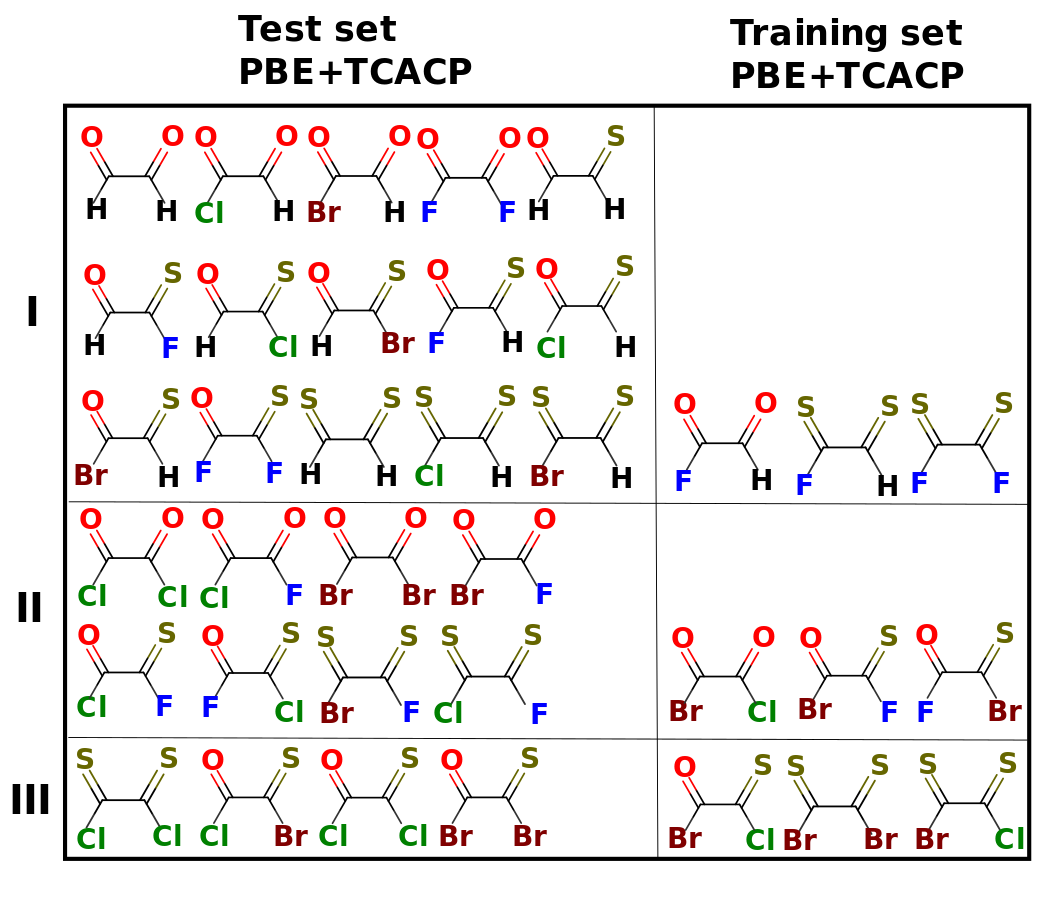}
\caption{\label{chemSpace} Overview of all molecules. Top, middle and bottom panels correspond to torsional profile classes displayed in Figure \ref{classes}. Three representative molecules from each class were used to parametrize torsion corrected atom-centered potentials (TCACP) shown in right column.}

\end{figure}

The choice of molecules limits our test to valence-only correlation as there are no core-valence polarized basis sets (cc-pCVXZ \cite{woon1995gaussian} or cc-pwCVXZ \cite{peterson2002accurate}) available for Br, which would recover core-valence correlation effects reliably. We have still considered two different valence-correlation schemes, one (A) correlating only valence \textit{s} and \textit{p} electrons (default in Molpro), the other (B) additionally considering 3\textit{d}-electrons of Br (default in Gaussian09 \cite{g09}). Results obtained with standard correlation-consistent basis sets \cite{dunning1989gaussian,woon1993gaussian,kendall1992electron} are collected in Table \ref{ccsdt_converge} for the energy difference $E(80^{\circ}) - E(180^{\circ}$).
\\ Regular cc-pVXZ basis sets obviously show erratic results for X=D, but fairly smooth convergence from X=T onwards, towards extrapolated \cite{ref_dirk} values listed in the table for both correlation schemes (A) and (B). Diffuse-augmentation generally leads to improved results for small basis but to very similar complete basis set (CBS) estimates. Explicitly correlated calculations at the CCSD(T)-F12 level with a triple-zeta basis set finally confirm large basis set and CBS values.\\
\\The fairly smooth convergence of aug-cc-pVXZ \cite{kendall1992electron} for correlation scheme (A) has prompted us to employ valence-\textit{sp}-correlated CCSD(T) with the relatively small aug-cc-pVTZ basis set (also referred to as AVTZ below) as a reference standard for all molecules included in this study (See Fig. \ref{chemSpace}).

\subsection{DFT results}

\begin{figure}[t]
\includegraphics[scale=0.27]{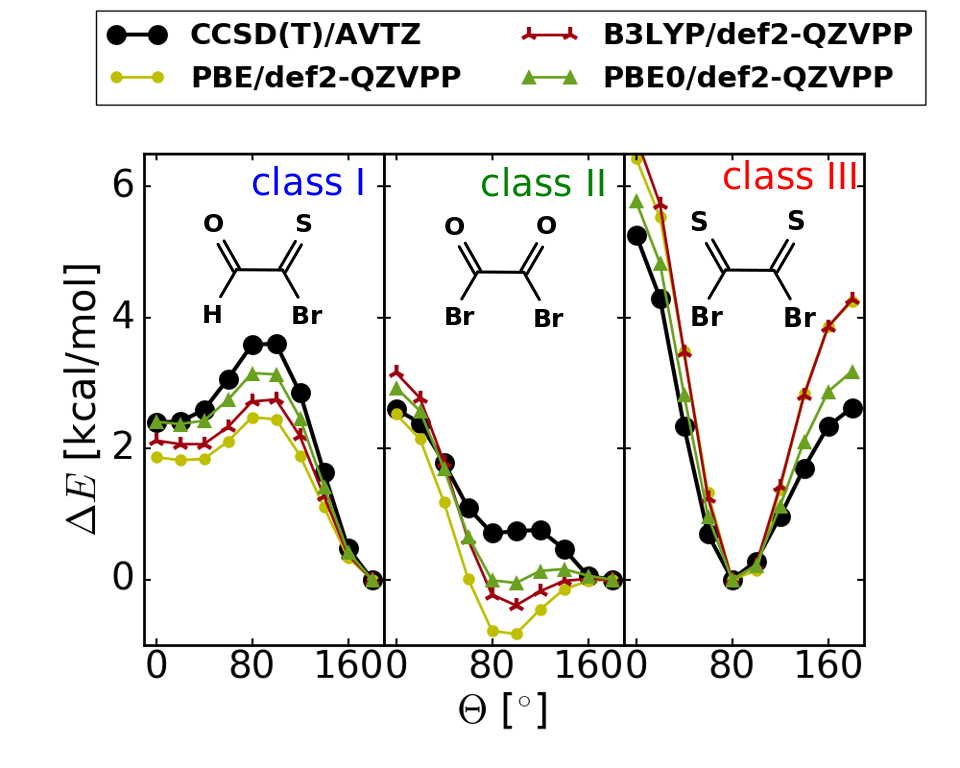}
\centering
\caption{\label{classes} Classification of torsional potentials based on their shape as predicted by CCSD(T). Specific CCSD(T) and DFT results are given for one representative molecule per class.}
\end{figure}

\begin{figure}[h]
\centering
\includegraphics[scale=0.23]{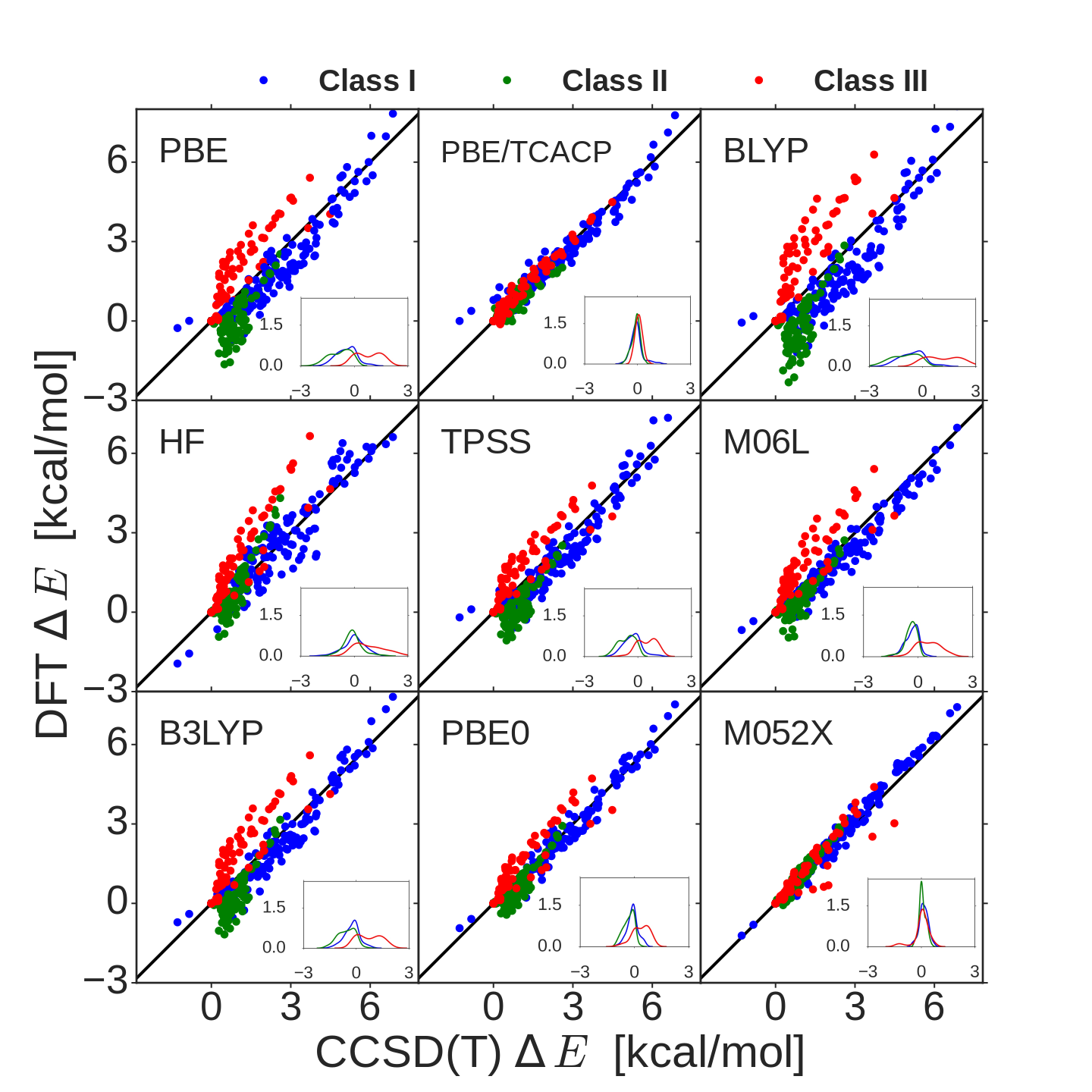}
\caption{ Scatter plots (DFT vs. CCSD(T)) of potential energies of all molecules in Fig. \ref{chemSpace} and for ten torsional angles, each relative to the energy for the conformation predicted to be the absolute minimum at the CCSD(T) level. Class I, II, and III profiles are shown in blue, green, and red, respectively. Correspondingly colored error distributions (DFT-CCSD(T)) are shown in insets. See Sec. 3.3 for a discussion of PBE+TCACP.}
\label{scatter_}
\end{figure}

\subsubsection{Overall performance}

The reliability of HF and popular DFT functionals was benchmarked against CCSD(T) results for the description of torsional potentials of 36 glyoxal, thiocarbonyl and (methanethioyl)-formaldehyde halides (Fig. \ref{chemSpace}). Torsional potentials at the CCSD(T) level may be grouped into three classes (Figs. \ref{chemSpace} and \ref{classes}). Systems with light halogens or hydrogen exhibit the conventional minima for $cis$ and $trans$ conformations that are stabilized by $\pi$-conjugation (class I). Substitution with heavier halogens with larger atomic radii introduces steric repulsion \cite{kim2005extremely}, which can only be relieved in non-planar conformations. We observe cases with only one minimum for orthogonal conformations (class III) and intermediate cases where only the $cis$ conformation becomes a transition state (class II). Here the $trans$ conformation remains the global minimum and it is typically augmented by a very shallow minimum for nearly orthogonal conformations. Class III compounds always contain at least one sulfur atom as well as chlorine and/or bromine. 
The very common GGA (PBE, BLYP) and hybrid (B3LYP, PBE0) density functionals are in qualitative agreement with CCSD(T) for classes I and III, but fail to reproduce the shape of class II torsional profiles (Fig. \ref{classes}).
\\Before scrutinizing differences between DFT and CCSD(T) one should note that any form of statistical analysis will crucially depend on the choice of reference point for the energy. One may shift DFT and CCSD(T) torsional profiles relative to each other such that the root mean squared deviation between DFT and CCSD(T) energies, sampled along the complete reaction coordinate, is minimized. This would allow for a ranking of functionals in terms of their overall accuracy.
\\Here we have deliberately chosen a different approach, namely to take that geometry of a molecule as a reference that turns out to be the global minimum at the CCSD(T) level. $\Delta E$ is then defined to be 0 for that geometry, not only at the CCSD(T) level but also at all DFT levels. This choice highlights problems of density functionals to account for the shape of a torsional profile and emphasizes errors in recovering torsional barriers.
\\Following this choice of reference, Figure \ref{scatter_} shows scatter plots of nine functionals vs. CCSD(T) for all molecules and angles (ten per molecule), as well as corresponding error distributions. Clearly most functionals overestimate the barriers of class III potentials (red) and underestimate them for class I potentials (blue). Their strong bias towards lower energies for class II potentials (green) indicates their tendency to overstabilize orthogonal geometries to the extent that these often become the global minimum (compare with Fig. \ref{classes}). Particularly serious problems are spotted for GGAs (PBE, BLYP), while the highly parametrized M05-2X performs well.

\begin{figure}[t]
\centering
\includegraphics[scale=0.42]{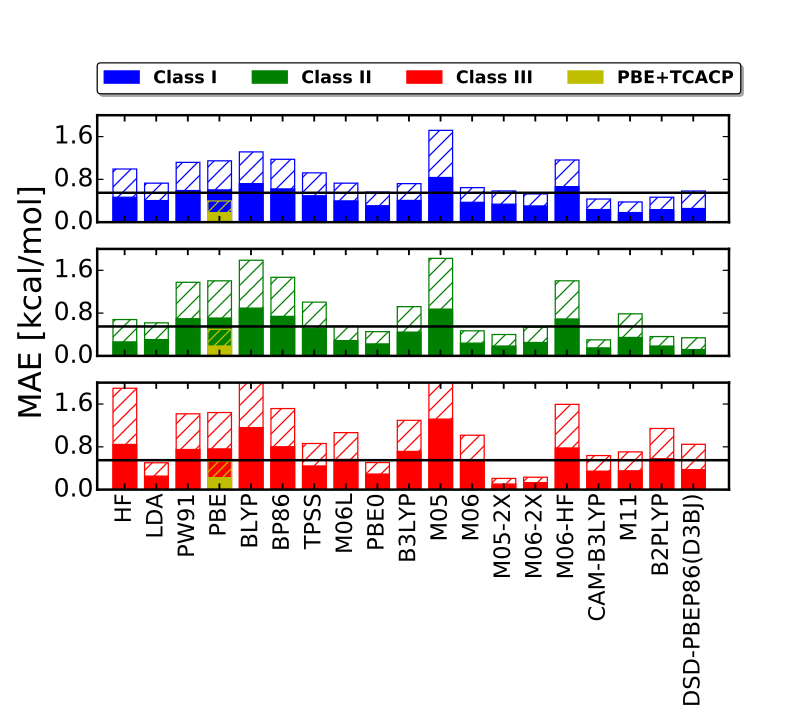}
\caption{ Mean absolute error (MAE, solid bars) and maximum absolute errors per molecule, averaged over all molecules of a given class (MaxAE, shaded bars). See caption of Fig. \ref{scatter_} further details. PBE+TCACP results (Sec. 3.3) are shown in yellow. Solid lines at 0.5 kcal/mol indicate errors that we consider acceptable for reasonable qualitative accuracy.}
\label{mae_}
\end{figure}

 \begin{figure}[t]
 \centering
 \includegraphics[scale=0.45]{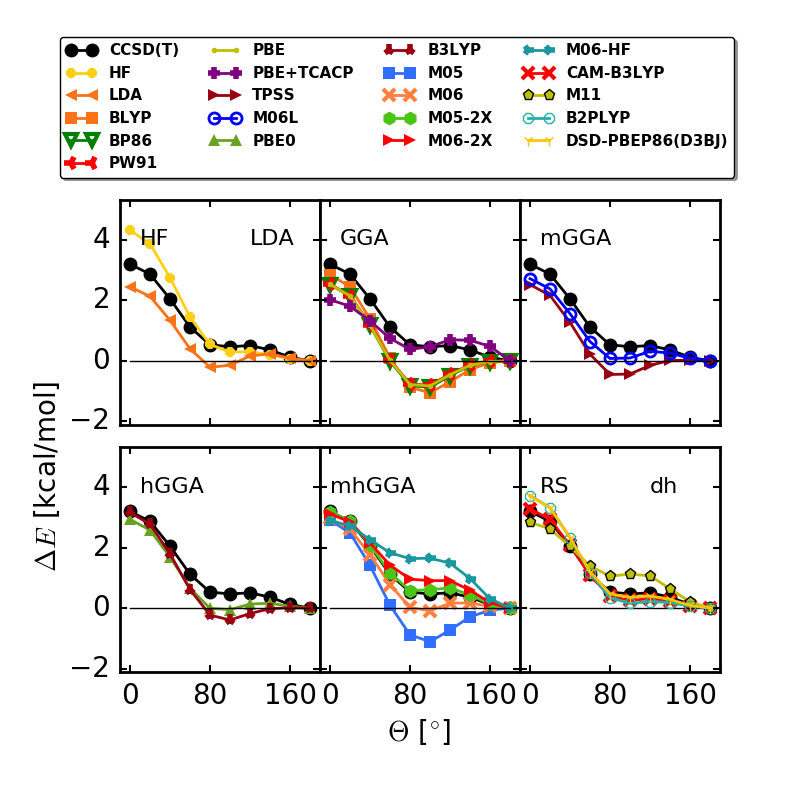}
 \caption{\label{bad_profile_worst} Torsional profile of oxalyl bromide (class II) obtained with popular density functionals and compared to CCSD(T) (in black).}
 \end{figure}

For the same choice of reference point, Figure \ref{mae_} provides a statistical analysis of HF and DFT energies compared to CCSD(T). Solid color bars illustrate the error averaged over all conformations of all molecules within the corresponding class of torsional profile (MAE, mean absolute error). Shaded color bars indicate the maximum error, found for any conformation of a molecule and averaged over all molecules within a class (MaxAE).
\\The small energy range of torsional potentials certainly requires an accuracy better than 0.5 kcal/mol in order to ensure a correct description of torsion. LDA shows a surprisingly small MAE of about 0.3 kcal/mol. One would hope for improvement upon climbing "Jacob's ladder \cite{perdew2001jacob}", however, the reverse is true. GGAs clearly perform worse than LDA, and the meta-hybrid M05 shows even larger errors. Large MaxAE values are particularly worrisome for class II profiles, they indicate spurious exaggeration of energy minima at orthogonal confirmation and thus a qualitatively wrong description. The performance of some functionals is still quite satisfactory for particular types of torsional profiles: M052-2X and M06-2X behave well for class III, \textcolor{black}{CAM-B3LYP and the double hybrids (B2PLYP and DSD-PBEP86(D3BJ)) are particularly successful for class I and II.} However there seems to be no functional that is universally recommendable for all three types of torsional potential. \\
Tables 1-3 of Supporting Information 2 list individual energy differences ($\Delta E$) between angles that refer to minima and maxima of the reference CCSD(T) profiles. Mean absolute errors with respect to CCSD(T) are listed as well and largely confirm the main conclusions drawn above: M05-2X and M06-2X reproduce barrier heights of class III molecules very well. CAM-B3LYP falls behind for class III, but is the best performer for classes I and II. 

\subsubsection{Oxalyl bromide}
Having assessed the general performance of DFT methods, we now scrutinize results for oxalyl bromide (Fig. \ref{bad_profile_worst}) as class II representative. CCSD(T) shows a very flat surface between $\Theta = 80^{\circ}$ and $\Theta = 140^{\circ}$. Close inspection indicates a very shallow minimum around $\Theta = 80^{\circ}$ and an equally shallow transition state at around $\Theta = 120^{\circ}$. Unconstrained optimization at M05-2X/def2QZVPP affords a minimum at $\Theta = 87.36^{\circ}$ and a maximum at $\Theta = 112.93^{\circ}$. Subsequent force constant analysis confirms the stationary points to be true a minimum and a true transition state, respectively. The corresponding barrier is very small, however (0.153 kcal/mol), and reduced further at the CCSD(T)/aug-cc-pVTZ//M05-2X/def2QZVPP level (0.04 kcal/mol). A definitive judgment on the existence of these intermediate points is therefore not attainable \cite{min_ts}.
\\Hartree-Fock behaves reasonably for the overall shape of the potential, but overestimates the $cis$ to $trans$ barrier; it shows a very flat potential between $\Theta=80^{\circ}$ and $140^{\circ}$ without extremal points. Many density functionals, however, predict deep and often global minima for perpendicular geometries. This is observed for LDA as well as for several popular density functionals of various rungs on Jacob's ladder, including mGGAs (TPSS), hGGAs (PBE0, B3LYP), and mhGGAs (M06, M06-HF). The GGAs (PBE, BLYP, PW91, BP86) and the meta-hybrid M05 functionals perform particularly poorly as they fail to reproduce the $trans$-minimum ($\Theta=180^{\circ}$). M06L is the best among the non-hybrid functionals, but only meta-hybrid (M05-2X, M06-2X), range-separated (CAM-B3LYP), and double hybrid (B2PLYP, \textcolor{black}{DSD-PBEP86(D3BJ)}) functionals manage to reproduce the finer details of the CCSD(T) reference. Similar remarks about relative performance hold for all molecules of class II and III as well as some cases of class I (all torsional profiles are given in Supporting Information 1).
\\One possible explanation for the failure of GGA functionals would be the presence of a small HOMO-LUMO gap at $\Theta \sim 80^{\circ}$, responsible for spurious charge transfer into the lowest unoccupied molecular orbital. However, we calculated a sizable gap of $\sim$ 3 eV ruling out this initial suspicion.

 \begin{figure}[tp]
 \includegraphics[scale=0.35]{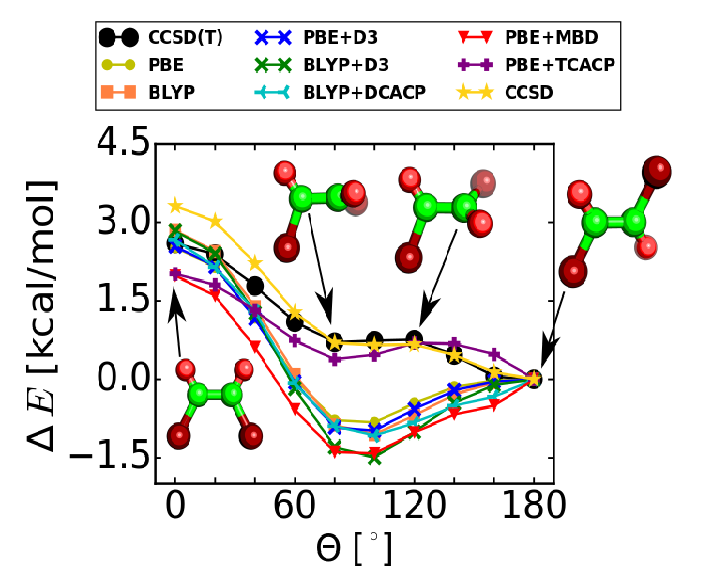}
 \caption{\label{corrections} Torsional profile of oxalyl bromide calculated with GGA functionals (PBE, BLYP), with and without TCACP or dispersion corrections (D3, DCACP and MBD), CCSD and CCSD(T).  }
 \end{figure}

\begin{table*}[htp!]
\caption{TCACP parameters [a.u.] for use in PBE functional}
\begin{tabular}{ l c c c c c c r }
 \multicolumn{7}{c}{} \\\hline
   & C & O & S & F & Cl & Br  \\\hline
 $r_{l=3}$ & 3.13  & 1.58 & 1.82 & 0.54 & 0.40 & 2.41 \\
 $h^{l=3}_{11}$ & $5.41\times10^{-4}$  & $-8.06\times10^{-3}$ & $-6.95\times10^{-3}$  & $7.04\times10^{-4}$  & $1.94\times10^{-3}$  & $-2.18\times10^{-3}$  \\\hline
\end{tabular}
\label{params}
\end{table*}

Considering that the interatomic separation between the two Br atoms in oxalyl bromide ranges from 3.3\:\AA\: to 4.7\:\AA \: in different conformations, one may further suspect a significant contribution of intramolecular dispersion interactions to the shape of torsional profiles. Figure \ref{corrections} shows torsional profiles calculated using popular dispersion corrections, namely: D3 \cite{grimme2010consistent}, Many Body Dispersion (MBD)\cite{tkatchenko2012accurate} and Dispersion Corrected Atom Centered Potentials (DCACP) \cite{von2004optimization,lin2007library}. None of the corrections improve the results of standard GGAs. {\color{black} In fact, they all increase the deviation from CCSD(T). (See also Table 4 in Supporting Information 2). Similar observations have been reported for intramolecular effects in a number of conformers and chemical reactions \cite{goerigk2014dft}.}

Furthermore, inspection of D3 dispersion corrections reveals maximal contributions for $80^{\circ} \leq \Theta \leq 100^{\circ}$, for which the Br-Br distance approaches the typical van der Waals minimum ($r_{\rm{Br-Br}} \approx 3.7$\AA). This explains the deepening of the spurious minimum. Consequently, lack of dispersion does not cause the observed problems, it rather lessens them for the wrong reason.
\\The real culprit instead appears to be the delocalization error. A pragmatic approach to reduce it is to include exact or Hartree-Fock exchange (HFx). The optimal admixture of exact exchange is $\sim$50\% (also used in the best functionals M05-2X, M06-2X, Figs. \ref{scatter_} - \ref{bad_profile_worst}). This observation is in agreement with earlier reports for conjugated double bonds \cite{ sancho2003assessment, sutton2014accurate}. Upon increasing the exact exchange further, the accuracy for torsional barriers is impaired again (M06-HF, HF).  
We tested this further by defining a functional (PBE-2X) that adds 56\% of exact exchange to standard PBE. This functional uses the same amount of exact exchange as M05-2X does and Fig. \ref{bad_profile_worst} shows that it is almost as accurate as the latter in reproducing the reference CCSD(T) torsional profile of oxalyl bromide. Additional tests indicate that any reduction or increase of exact exchange worsens the results of PBE-2X (data not shown).

\subsection{TCACP corrections}

While some hybrid density functionals (M05-2X, M06-2X, PBE-2X) can provide satisfactory descriptions of torsional profiles, the calculation of exact exchange makes them computationally inefficient for use with plane-wave basis sets. PWs are predominantly employed in condensed phase studies where functionals at the GGA level are a common compromise between efficiency and accuracy. Improvements for GGA based predictions on atom centered corrections (typically implemented in the form of pseudo potentials) were found for various properties, such as London dispersion \cite{von2004optimization,von2005variational,lin2007library,dilabio2008accurate,prasad2018atom}, vibrational frequencies\cite{lilienfeld2013force}, band gaps \cite{christensen1984electronic,segev2007self,von2008structure} and relativistic effects \cite{bachelet1982relativistic,dolg2011relativistic}. Atom centered corrections have further been used to reduce basis set incompleteness effects \cite{otero2017transferable}. Therefore, we have studied if one can improve the PBE prediction of rotational profiles using custom generated torsion corrected atom centered potentials (TCACP). 

\subsubsection{Calibration}
\begin{figure}[t]
\centering
\includegraphics[scale=0.35]{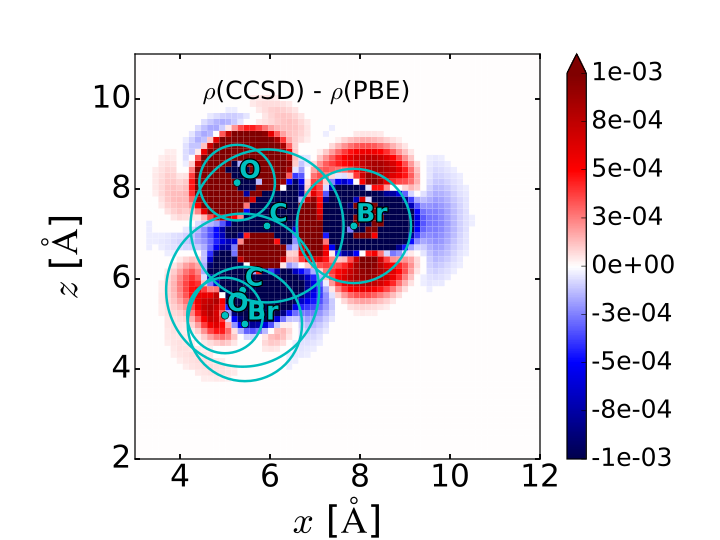}
\\
\includegraphics[scale=0.35]{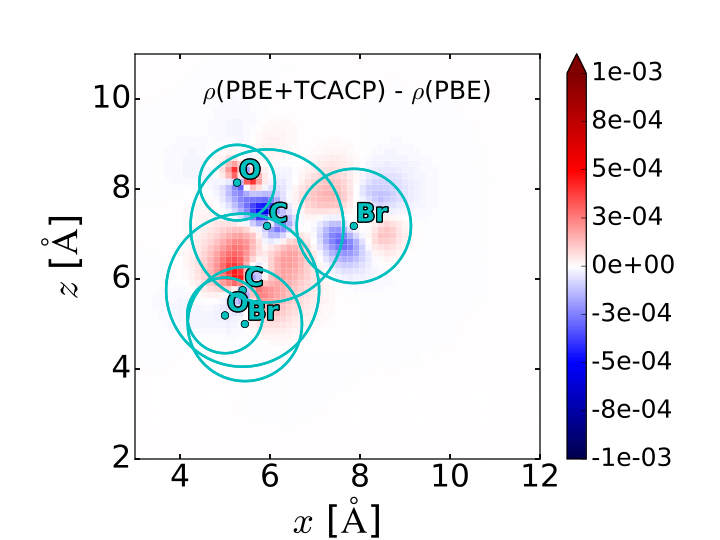}
\caption{\label{dens} Density difference plots for oxalyl bromide at $\Theta=80^{\circ}$: PBE versus CCSD (top) and versus PBE+TCACP (bottom). The range at which the TCACPs reach their maximum is illustrated by cyan circles. Isosurface density differences ($\pm0.0001$ a.u.) are given as insets.} 
\end{figure}

Torsion corrected effective potentials are optimized for C, O, S, F, Cl and Br (Table \ref{params}). As a training set we used 10 energies each along the torsional profiles of nine molecules, selected from all three classes (Fig. \ref{chemSpace}, right column). The optimized parameters (shown in Table \ref{params}) indicate distance $r_{\textit{l}}$ and magnitude $h_{\rm{11}}^{\textit{l}}$ at which the PBE+TCACP calculations reach maximum agreement with the reference. Note that TCACPs centered on Cl and F peak at $\sim$ 0.5 Bohr from the nuclei, whereas for larger atoms, such as Br, S and O, as well as C, $r_{\textit{l}}$ is larger than 1.5 Bohr (Fig. \ref{dens}, shown for O, Br, and C in cyan circles). Further note that corrections centered on Br have opposite sign compared to other halogens. A positive (negative) $h_{\rm{11}}^{\textit{l}}$ indicates that the original PBE exchange-correlation underestimates (overestimates) the reference energy.

\subsubsection{Electron density}
One might expect that TCACPs improve on density functional
descriptions through correction of electron densities. Analysis
of density differences, however, disproves this
hypothesis, at least for the case examined. Although
TCACPs successfully eliminate the fairly large energy error for
oxalyl bromide at $\Theta$ = 80$^{\circ}$ (Fig. \ref{corrections}), as discussed below, the density difference generated (Fig. \ref{dens}, bottom) is not only much smaller but also largely of opposite sign compared to the one calculated with CCSD as reference (Fig. \ref{dens}, top). We assume that the use of CCSD instead of CCSD(T) as reference is reasonable, as both methods produce very similar energy curves (Fig. \ref{corrections}). We must, therefore, conclude that TCACPs act directly through a change in the external potential, rather than indirectly through a mediated change of electron density.

\subsubsection{Geometry and frequencies}
For general use, it is important to show that TCACPs do not negatively affect other already well-described properties, such as equilibrium geometries and vibrational frequencies. Tests were performed for oxalyl bromide in $cis$, $gauche$, and $trans$ conformations (results not shown in detail). Addition of TCACPs changes bond lengths by less than 0.01 \AA \: and angles by less than 0.3$^{\circ}$. Vibrational frequencies are affected by 32 cm$^{-1}$ at most and 23 cm$^{-1}$ on average. Changes of this order are insignificant for most applications, suggesting that the addition of TCACPs to PBE does not lead to adverse side effects.


\begin{figure*}[t]
    \centering
        \includegraphics[scale=0.285]{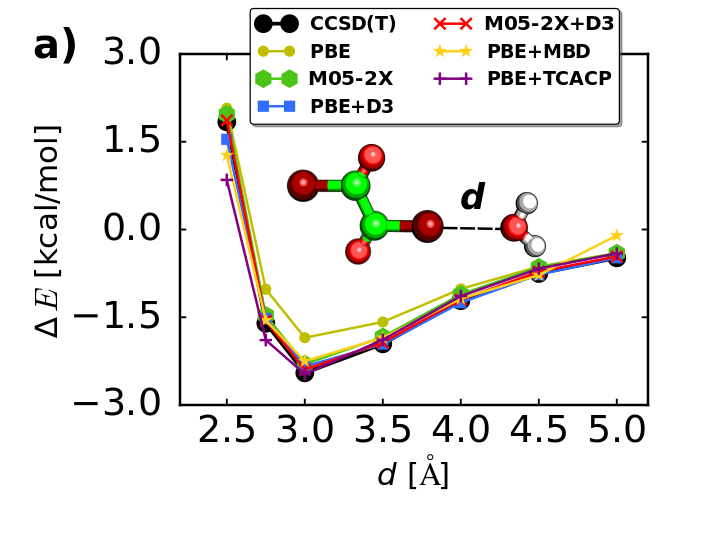}
       \label{b_en}
    \hfill
        \includegraphics[scale=0.285]{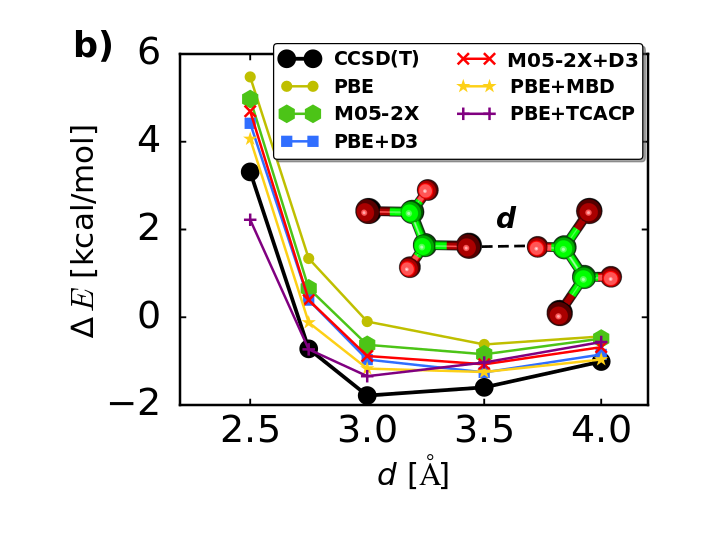}
     \label{dimer_en}
    \hfill
        \includegraphics[scale=0.2]{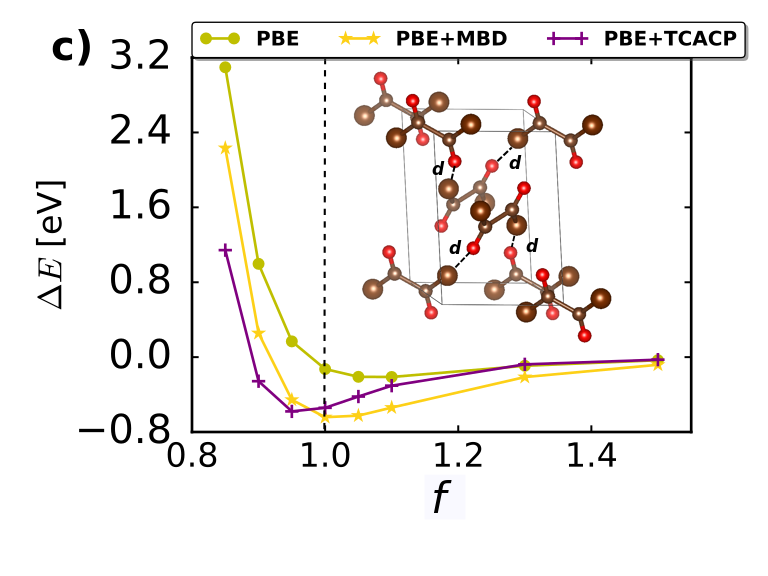}
    \label{crystal}
    \caption{ \label{bind_en} Interaction energy curves of oxalyl bromide with water a), dimer b), and in crystal structure c) obtained using PBE with and without corrections and compared to M052X and CCSD(T) (a, b). Experimental geometry \cite{groth1962crystal,damm1965dioxan} has been used in c), and energies are shown as a function of $f$, the ratio between assumed and experimental lattice constant.}
\end{figure*}

\subsubsection{Torsional profiles}

Results obtained from PBE+TCACP are included in Figs. \ref{scatter_}, \ref{mae_} and \ref{corrections}. They all show that TCACPs lead to the desired improvement in the description of rotational profiles. Figure \ref{corrections} illustrates for oxalyl bromide how TCACPs correct the qualitatively wrong profile and reach good agreement with CCSD(T). The effect is most pronounced for intermediate values of $\Theta$. The corrected functional PBE+TCACP still underestimates the barrier by about 1 kcal/mol, but it clearly shows the overall best performance of all GGA functionals. Note that our test case, oxalyl bromide, was not even included in the TCACP training set (Fig. \ref{chemSpace}).  Figure \ref{scatter_}, reporting correlations and error distributions, indicates that the corrections are transferable to all molecules included.
\\Further evidence for the overall good performance of TCACP corrections is obtained from Fig. \ref{mae_}: The relatively large PBE error is reduced to less than 0.5 kcal/mol (black line), which is the threshold we used to identify qualitatively satisfying results. MaxAE is now within the range of 0.5 kcal/mol for all types of torsional profile, which is below the MAE of any of the GGAs, and even lower than some of the hybrid and meta-hybrid functionals. Likewise the MAE for barrier heights (Supporting Information II, Tables 1-3) is among the lowest of all functionals for classes I and II and significantly smaller than PBE also for class III. The complete set of profiles given in the Supporting Information 1 corroborates the overall good performance among all molecules studied here.
\\All these observations suggest that TCACP corrections are a suitable empirical way to rectify some of PBE's shortcomings in modeling torsional profiles.

\subsubsection{Intermolecular interactions}
After studying the performance of TCACPs for torsional potentials, we were curious to assess their effect on intermolecular interactions.
To keep in line with our general choice of test cases, we are again considering oxalyl bromide, now a) in complex with water, b) as a dimer, and c) as a crystal (Fig. \ref{bind_en}), which are all thought to benefit from $\sigma$-hole-binding \cite{politzer2007sigma}. CCSD(T) was used as a binding energy reference in the first two cases but is computationally prohibitive for studying the crystal. No experimental data is available for the cohesive energy of the crystal, restricting us to a qualitative discussion based on the experimental geometry \cite{damm1965dioxan,groth1962crystal} only.
\\PBE underestimates binding energies in all three cases (Fig. \ref{bind_en}), and shows hardly any minimum for the oxalyl bromide dimer or the crystal. M05-2X definitely performs better for the complex with water but still falls short of expectations for the dimer. Crystal structure energy evaluation with M05-2X would be computationally too demanding and have thus not been attempted.
Dispersion corrections (MBD, D3) remedy the shortcomings of PBE for the water complex and reach the good performance of M05-2X. In the case of PBE, they also lead to an improved description of the dimer. They show little effect for M05-2X. The crystal structure, finally, benefits from dispersion corrections to PBE, as the PBE+MBD energy minimum is now obtained for the experimental lattice constant. 
\\Although parametrized for torsional potentials only, TCACPs also help to improve binding energy curves. They correct the underbinding of the water complex observed with PBE and improve on the description of the dimer and the crystal. Agreement with the CCSD(T) reference for the dimer is actually a little better for PBE+TCACP than for dispersion-corrected PBE, where the crystal lattice constant may be slightly underestimated. Overall both types of correction perform quite similarly. This might indicate that TCACP unintentionally picks up some of the dispersion missing in PBE, at least for the few cases studied here.

\section{Conclusions}

An extensive search in literature shows that the performance of density functional approximations for the description of torsional profiles have been mostly tested for systems with small average nuclear charges. Thus, the selection bias on how we test the methods slips cases where the DFT methods fail.
\\
The performance of popular density functional methods was assessed for the description of torsional profiles of 36 molecules: Glyoxal and oxalyl halides, and their thiocarbonyl derivatives. Reference calculations at the CCSD(T) level show that the choice of halogen determines the shape of the profile, and three distinct classes of profile have been identified.
\\
Most density functionals used in this study fail to reproduce barriers accurately and even show qualitatively incorrect profiles for molecules belonging to class II. The initial suspicion that spurious charge transfer might cause the problems could not be substantiated. Lack of dispersion was also ruled out because the tested dispersion corrections (D3 \cite{grimme2010consistent}, MBD \cite{tkatchenko2012accurate}, DCACP \cite{von2004optimization,lin2007library}) only worsened the predictions.
\\Further analysis shows that GGA as well as the meta-hybrid M05 functionals perform worst and that addition of about 50\% exact exchange cures most of the problems. Larger amounts of exact exchange tend to impair the accuracy again. The treatment of exchange interactions is thus identified as the core problem of DFT in reproducing torsional barriers of glyoxal derivatives substituted with sulfur and/or heavier halogens.
This explanation is consistent with the fact that PBE-2X results in a vastly improved rotational profiles for oxalyl bromide.
\\Inclusion of exact exchange, unfortunately, precludes DFT calculations with plane wave basis sets as they demand pure functionals for computational efficiency, particularly in large-scale materials applications. Torsion-corrected atom-centered potentials, TCACPs, have been found to provide a simple, empirical way out of this dilemma. With only 2 parameters per element, they improve the accuracy of standard PBE to a level comparable to the best hybrid functional (M05-2X). TCACPs work well also for molecules outside the training set, and they have little effect on other properties, such as optimized geometries, vibrational frequencies, and electron density distributions. Although parametrized for torsional potentials, TCACPs also improve intermolecular binding potentials, at least for the limited number of test cases studied so far.
\textcolor{black}{Thus the design of atom-centered potential corrections which target multiple flaws of GGAs may be a valuable goal in future studies.}

\section{Acknowledgment}
The authors thank S. Willitsch, D.~Alf\`e, M. Schwilk, and A.~Tkatchenko for suggesting cis-trans isomerism, Quantum Monte Carlo calculations, CCSD(T)-F12 calculations, and discussions, respectively.
O.A.v.L. acknowledges funding from the Swiss National Science foundation (No.~PP00P2\_138932, 310030\_160067). This research was partly supported by the NCCR MARVEL, funded by the Swiss National Science Foundation. Some calculations were performed at sciCORE (http://scicore.unibas.ch/) scientific computing core facility at University of Basel. 

\section{SI}

Supporting Information document 1 provides figures with torsional profiles for all molecules and all levels of theoty considered in this work. \\
Supporting Information document 2 contains additional tables with (i) individual energy differences ($\Delta E$) between angles that refer to minima and maxima of the reference CCSD(T) profiles as well as MAEs with respect to CCSD(T); (ii)  data used to generate Figure 7.\\
Supporting information document 3 provides all optimized geometries and energies in computer readable format {\tt geometries.txt}.

The Supporting Information is available free of charge on the ACS Publications website.

\bibliographystyle{apsrev4-1}
\bibliography{ref}

\clearpage
\begin{figure*}
\centering
\includegraphics[scale=0.7]{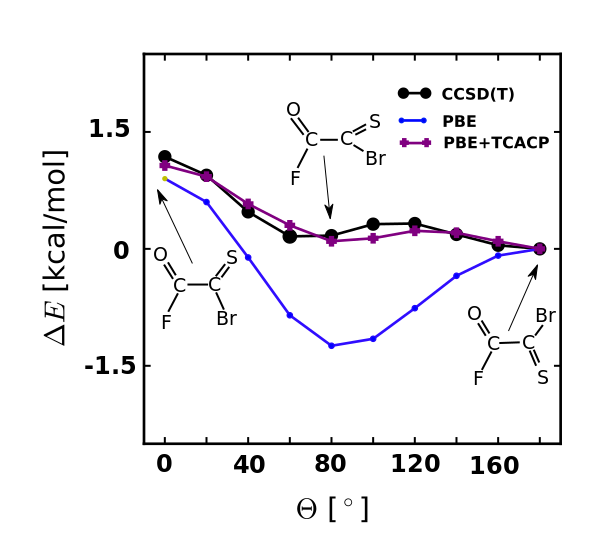}
\caption{TOC}
\end{figure*}

\end{document}